\documentclass[aps, prb, showpacs, preprint, amsmath, letterpaper]{revtex4-1}

\usepackage{graphics}
\usepackage{graphicx}
\usepackage{amssymb}
\usepackage{bm}

\begin{document}
\title{Extreme magnetic field-boosted superconductivity}

\author{Sheng Ran$^{1,2,3*}$, I-Lin Liu$^{1,2,3}$, Yun Suk Eo$^{1}$, Daniel J. Campbell$^{1}$, Paul Neves$^{1}$, Wesley T. Fuhrman$^{1}$, Shanta R. Saha$^{1,2}$, Christopher Eckberg$^{1}$, Hyunsoo Kim$^{1}$, Johnpierre Paglione$^{1,2}$, David Graf$^{4}$, John Singleton$^{5,6}$ \& Nicholas P. Butch$^{1,2,*}$}

\affiliation{$^1$ Center for Nanophysics and Advanced Materials, Department of Physics, University of Maryland, College Park, MD 20742, USA
\\$^2$ NIST Center for Neutron Research, National Institute of Standards and Technology, Gaithersburg, MD 20899, USA
\\$^3$ Department of Materials Science and Engineering, University of Maryland, College Park, MD 20742, USA
\\$^4$ National High Magnetic Field Laboratory, Florida State University, Tallahassee, FL 32313, USA
\\$^5$ National High Magnetic Field Laboratory, Los Alamos National Laboratory, Los Alamos, NM 87545, USA
\\$^6$ Department of Physics, The Clarendon Laboratory, University of Oxford, Oxford OX12JD, United Kingdom
}

\date{\today}

\begin{abstract}
Applied magnetic fields underlie exotic quantum states, such as the fractional quantum Hall effect~\cite{Stormer1999} and Bose-Einstein condensation of spin excitations~\cite{Zapf2014}. Superconductivity, on the other hand, is inherently antagonistic towards magnetic fields. Only in rare cases~\cite{Meul1984,Uji2001,Konoike2004} can these effects be mitigated over limited fields, leading to reentrant superconductivity. Here, we report the unprecedented coexistence of multiple high-field reentrant superconducting phases in the spin-triplet superconductor UTe$_{2}$~\cite{Ran2019}. Strikingly, we observe superconductivity in the highest magnetic field range identified for any reentrant superconductor, beyond 65~T. These extreme properties reflect a new kind of exotic superconductivity rooted in magnetic fluctuations~\cite{Mineev2015} and boosted by a quantum dimensional crossover~\cite{Lebed2014}.




\end{abstract}
\maketitle
It is a basic fact that magnetic fields are destructive to superconductivity. The maximum magnetic field in which superconductivity survives, the upper critical field $H_{c2}$, is restricted by both the paramagnetic and orbital pair-breaking effects: electron spin polarization due to the Zeeman effect, and electron cyclotron motion due to the Lorentz force, respectively. In a few very rare cases, however, magnetic fields can do the opposite and actually stabilize superconductivity~\cite{Meul1984,Uji2001,Konoike2004}. In these cases, the applied magnetic field is most often compensated by an internal field produced by ordered magnetic moments through exchange interactions, resulting in a reduced total effective field~\cite{Jaccarino1962}. A different set of circumstances involving unconventional superconductivity occurs in the ferromagnetic superconductor URhGe~\cite{Levy2005,Levy2007}, where field-induced superconductivity is instead attributed to very strong ferromagnetic fluctuations that emanate from a quantum instability of a ferromagnetic phase, strengthening spin triplet pairing~\cite{Mineev2015}.

\begin{figure}
\includegraphics[angle=0,width=180mm]{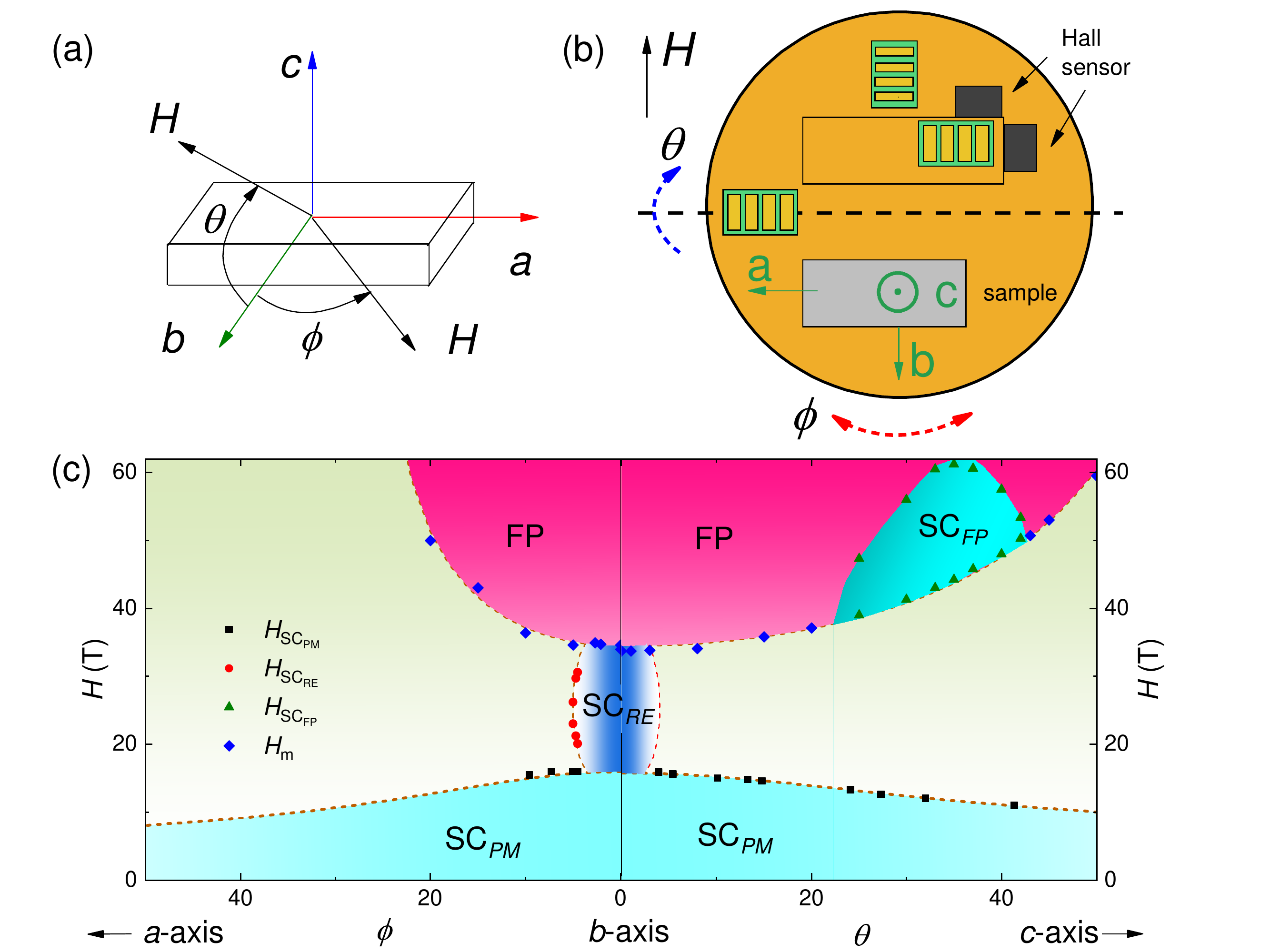}
\caption{Magnetic field-induced superconducting and polarized phases of UTe$_{2}$. (a) Sketch of how the magnetic field is applied with respect to the three crystallographic axes of UTe$_{2}$. (b) Top view of the sample platform with two-axis rotator used in DC field measurements to achieve the best alignment. (c) Magnetic field - angle phase diagram showing three superconducting phases. The magnetic field is rotated within the $ab$ and $bc$-plane. The critical field values of the SC$_{\rm PM}$ and SC$_{\rm RE}$ phases are based on DC field measurements, and those of the SC$_{\rm FP}$ and field polarized phases are based on pulsed field measurements.}
\label{2DanglePD}
\end{figure}

Here we report the presence of two independent high-field superconducting phases in the recently discovered triplet superconductor UTe$_{2}$~\cite{Ran2019}, for a total of three superconducting phases (Fig.~\ref{2DanglePD}). This is the first example of two field-induced superconducting phases existing in one system, one of which has by far the highest lower and upper limiting fields of any field-induced superconducting phase, more than 40~T and 65~T, respectively. Both of the field-induced superconducting phases are stabilized by ferromagnetic fluctuations that are induced when the magnetic field is applied perpendicular to the preferred direction of the electron spins. Interestingly, the high-field superconducting phase exists exclusively in a magnetic field polarized state, unique among these superconductors. This discovery provides an excellent platform to study the relation between ferromagnetic fluctuations, spin triplet superconducting pairing, and dimensionality in the quantum limit.

\begin{figure}
\includegraphics[angle=0,width=175mm]{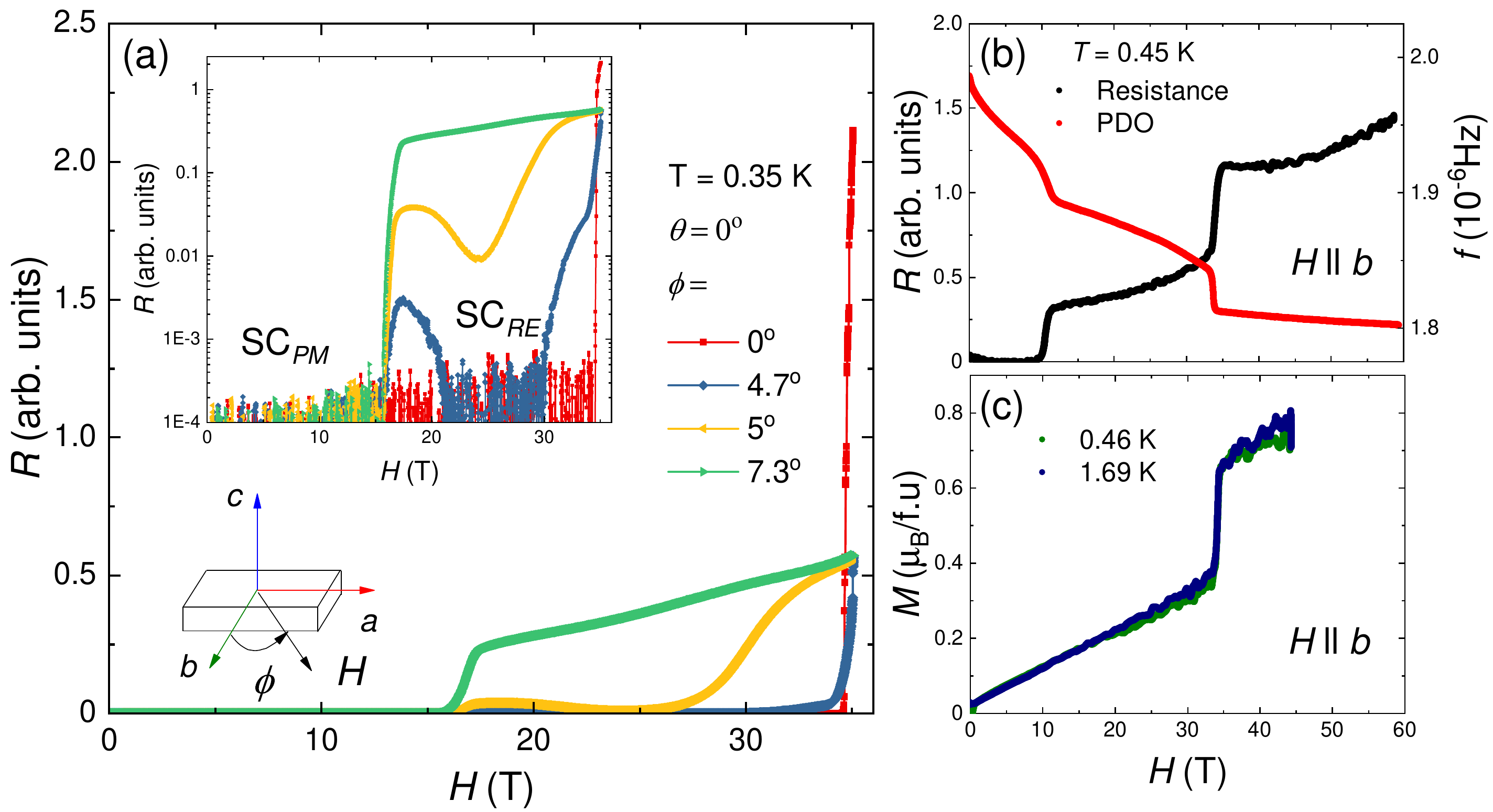}
\caption{Reentrance of superconductivity in UTe$_{2}$. (a) Field dependence of magnetoresistance of UTe$_{2}$ at $T$ = 0.35~K measured in the DC field. The magnetic field is rotated from $b$ towards $a$ axis. Zero resistance persists up to 34.5~T when the magnetic field is perfectly along $b$-axis. The same data set is plotted in logarithmic scale in the inset. Reentrance of superconductivity can be clearly seen when the magnetic field is applied slightly off the $b$-axis. (b) Magnetoresistance and the contactless-conductivity measurements using the PDO circuit (see Methods section for the technical details) of UTe$_{2}$ at $T$ = 0.45~K in the pulsed field, with the magnetic field applied along the $b$-axis. (c) Magnetization measurements UTe$_{2}$ at $T$ = 0.45~K and 1.7~K in the pulsed field, with the magnetic field applied along the $b$-axis. For the measurements in pulsed field, the two-axis rotator is not compatible. There is likely a slight angle offset along the perpendicular direction. The field induced superconducting phase SC$_{\rm RE}$ is not observed in these measurements.}
\label{Baxis}
\end{figure}

UTe$_{2}$ crystallizes in an anisotropic orthorhombic structure, with the $a$-axis being the magnetic easy axis along which spins prefer to align in low magnetic fields. The superconducting upper critical field $H_{c2}$ is strongly direction-dependent, and exceedingly large along the $b$-axis, with an unusual upturn in its temperature dependence above 15~T. The $H_{c2}$ is extraordinarily sensitive to the alignment of the magnetic field along the $b$-axis~\cite{Aoki2019}, and accurate measurements require the use of a specialized two-axis rotator (Fig.~\ref{2DanglePD}b). When the magnetic field is perfectly aligned along the $b$-axis, superconductivity persists up to 34.5~T at 0.35~K (Fig.~\ref{Baxis}a). A small misalignment from the $b$-axis, less than 5$^{\circ}$, decreases the $H_{c2}$ value by over half, to 15.8~T. However, even this misaligned superconductivity is resilient, and upon further increasing the magnetic field, superconductivity reappears between 21~T and 30~T. Our measurements show that this reentrant phase does not persist beyond misalignment greater than 7$^{\circ}$.


Although UTe$_{2}$ is closely related to the ferromagnetic triplet superconductors~\cite{Ran2019}, the observation of reentrant superconductivity in UTe$_{2}$ resembles neither that of URhGe~\cite{Levy2005,Levy2007}, which is completely separated from the low-field portion, nor the sharp $H_{c2}$ cusp in UCoGe in angle dependence~\cite{Aoki2009}. The angle-dependence of the superconducting phase boundary suggests that the reentrant portion of the UTe$_{2}$ superconductivity, SC$_{\rm RE}$, may have a distinct order parameter from the lower-field superconductivity, SC$_{\rm PM}$. However, unlike the case for both URhGe~\cite{Levy2007,Huxley2007} and UCoGe~\cite{Aoki2009,Hattori2014}, there is no normal-state change in the underlying magnetic order in UTe$_{2}$ that would drive a change in the superconducting order parameter symmetry. We discuss the magnetic interactions that stabilize this unusual behavior after an excursion to even higher field.


The upper field limit of SC$_{\rm RE}$ of 35~T coincides with a dramatic magnetic transition into a field polarized phase (Fig.~\ref{Baxis}c). The magnetic moment along the $b$-axis jumps from 0.35 to 0.65~$\mu_{B}$ discontinuously, due to a spin rotation from the easy $a$-axis to the orthogonal $b$-axis. The abrupt change in moment direction is accompanied by a jump in magnetoresistance and a sudden change of frequency in the proximity detector oscillator (PDO) circuit (Fig.~\ref{Baxis}b). The critical field $H_m$ of this magnetic transition has little temperature dependence up to 10~K, but increases as magnetic field rotates away from the $b$-axis to either the $a$ or $c$-axis (Fig.~\ref{2DanglePD}c). Meanwhile, the magnitude of the jump in magnetic moment, 0.3~$\mu_{B}$, appears to be direction-independent (Supplemental material Fig. 5). This magnetic field scale appears to represent a general energy scale for correlated uranium compounds: weak anomalies are observed in ferromagnetic superconductor UCoGe~\cite{Knafo2012}, whereas a large magnetization jump occurs in the hidden order compound URu$_{2}$Si$_{2}$~\cite{DeBoer1986}.



\begin{figure}
\includegraphics[angle=0,width=180mm]{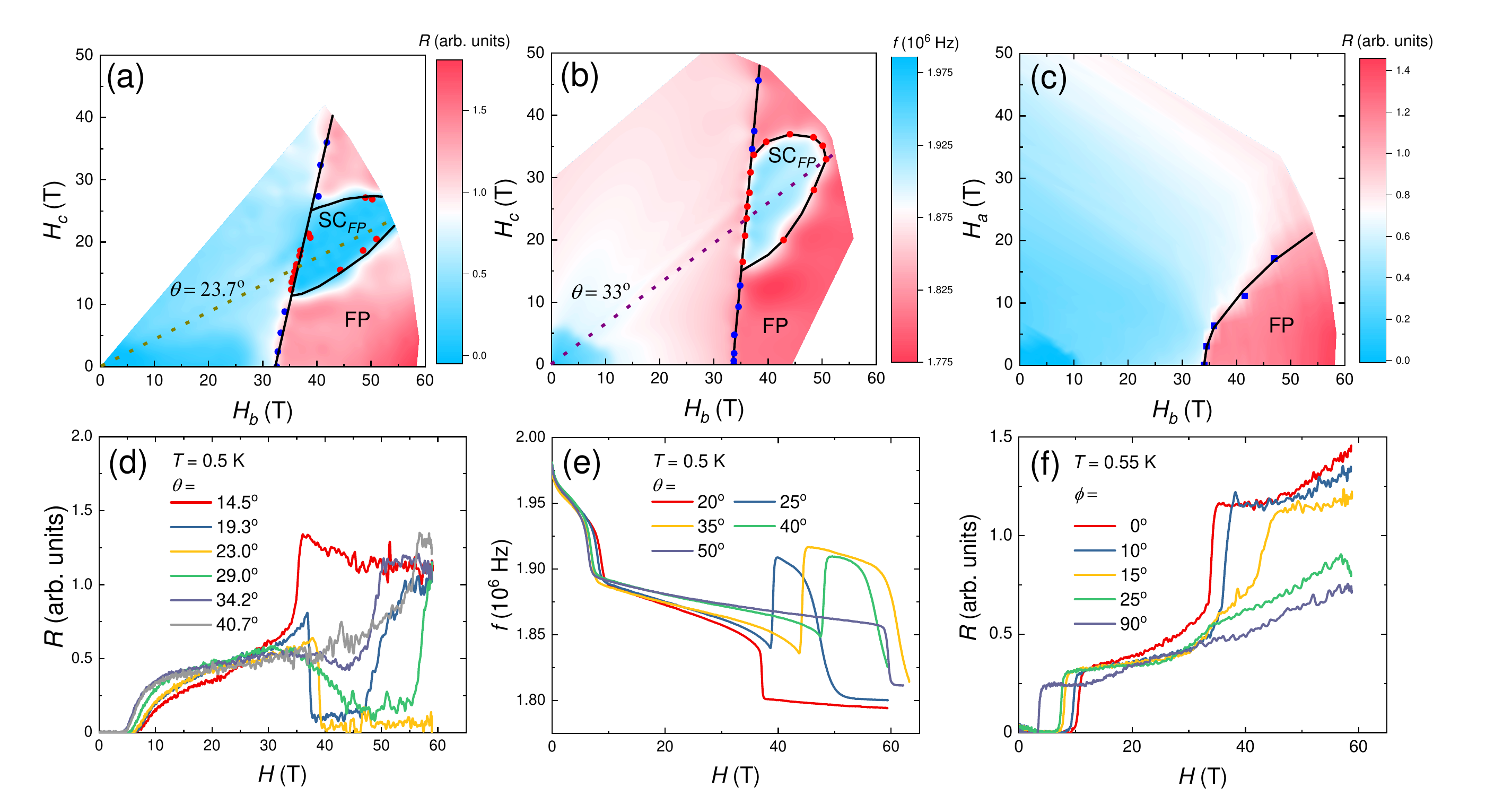} 
\caption{Angle dependence of the field-induced superconducting and polarized phases of UTe$_{2}$. When the magnetic field is applied at an angle from $b$ towards $c$-axis ($a$-axis), it is equivalent to two applied magnetic fields: one along the $b$-axis, $H_b$ = $H$cos$\theta$ or $H$cos$\phi$, and the other along the $c$-axis, $H_c$ = $H$sin$\theta$ ($a$-axis, $H_a$ = $H$sin$\phi$). It is more convenient to visualize the magnetoresistance and PDO data as a function of these two components of the applied magnetic fields. Such color contour plots are shown in (a) - (c). The blue dots are the critical fields for the field polarized state and the red dots are the critical fields for the high-field superconducting phase SC$_{\rm FP}$. The corresponding data as a function of the applied magnetic fields are shown in (d) - (f) at selected angles. The dashed lines on the (a) and (b) indicates the directions where measurements were also performed at different temperatures, as shown in next figure.} 
\label{Reentrance}
\end{figure}

As $H_{m}$ limits the SC$_{\rm RE}$ phase, it gives rise to an even more startling form of superconductivity. Sweeping magnetic fields through the angular range of $\theta$ = 20 - 40$^{\circ}$ from the $b$ towards the $c$-axis reveals an unprecedented high-field superconducting phase, SC$_{\rm FP}$ (Fig.~\ref{Reentrance}). The onset field of the SC$_{FP}$ phase precisely follows the angle dependence of $H_{m}$, while the upper critical field goes through a dome, with the maximum value exceeding 65~T, the maximum field possible in our measurements. This new superconducting phase significantly exceeds the magnetic field range of all known field-induced superconductors~\cite{Meul1984,Uji2001,Konoike2004,Levy2005}. Due to its shared phase boundary with the magnetic transition, this superconducting phase tolerates a rather large angular range of offsets from the $b$-$c$ rotation plane. However, it does not appear when the field is rotated from the $b$- to the $a$-axis.

\begin{figure}
\includegraphics[angle=0,width=175mm]{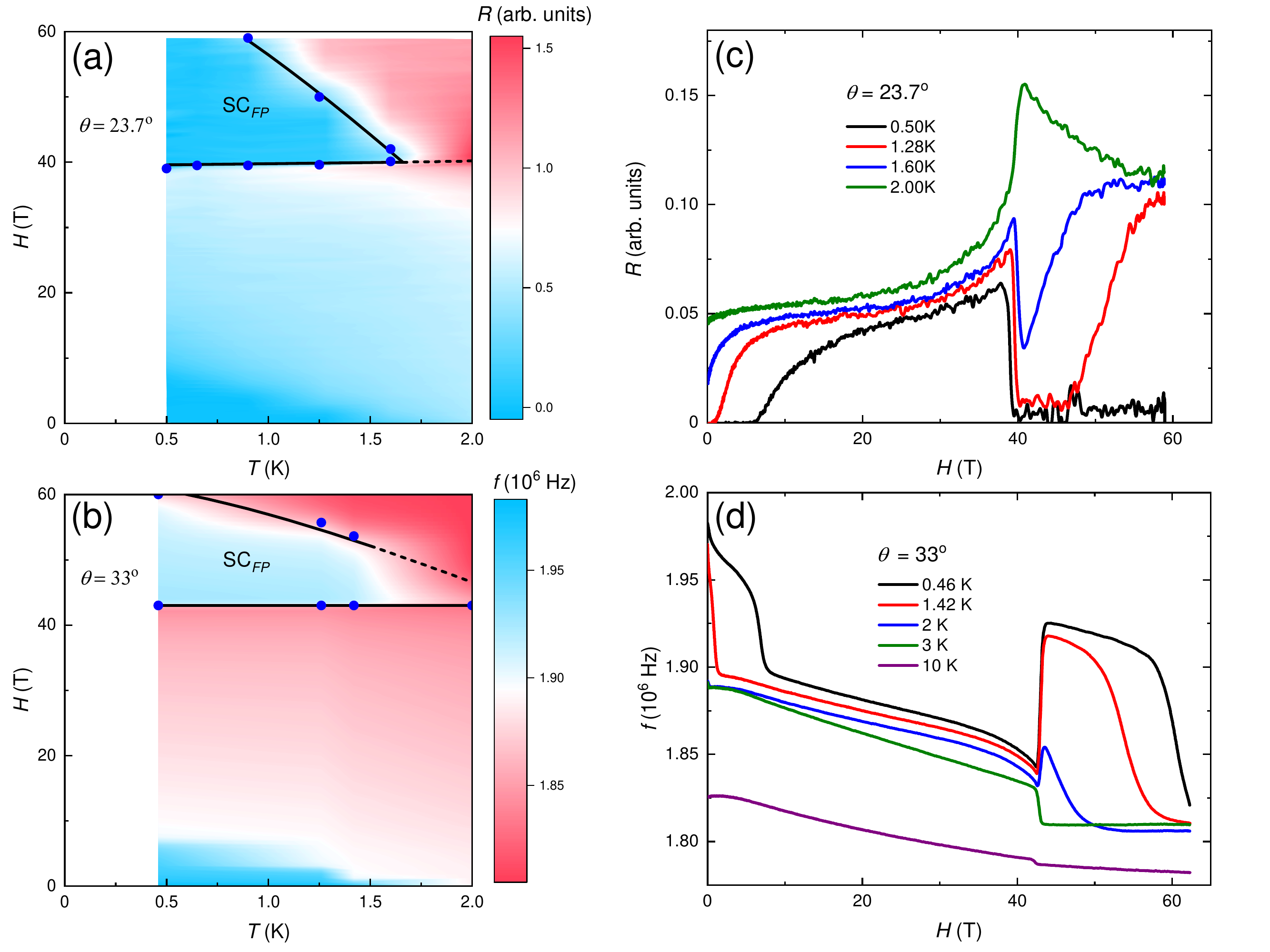} 
\caption{Temperature dependence of the high-field superconducting phase SC$_{\rm FP}$ of UTe$_{2}$. Color contour plots of (a) magnetoresistance and (b) frequency of PDO measurements as a function of temperature and magnetic field, at $\theta$ = 23.7$^{\circ}$ and 33$^{\circ}$, respectively. The corresponding data as a function of the applied magnetic fields are shown in (c) and (d) at selected temperatures.} 
\label{RESCT}
\end{figure}

Having established the field limits and angle dependence of the SC$_{\rm FP}$ phase, we turn to its temperature stability (Fig.~\ref{RESCT}). The onset field has almost no temperature dependence, again following $H_{m}$, while the upper critical field of the SC$_{\rm FP}$ phase disappears near 1.6~K, similar to the zero-field superconducting critical temperature. This suggests that even though it is stabilized at remarkably high field, the new superconducting phase involves a similar pairing energy scale to the zero-field superconductor. 

Such a large magnetic field and temperature stability of the SC$_{\rm FP}$ phase begs the question of what the mechanism is. A natural candidate is the Jaccarino-Peter effect used to describe other reentrant superconductors~\cite{Jaccarino1962}. This antiferromagnetic type of exchange interaction can lead to an internal magnetic field that is opposite to the external magnetic field, resulting in a total magnetic field that is much smaller. This compensation mechanism has successfully explained the field induced superconductivity in Chevrel phase compounds and organic superconductors~\cite{Meul1984,Uji2001,Konoike2004}, but it is very unlikely to apply to the SC$_{\rm FP}$ phase of UTe$_{2}$, which lacks the requisite localized atomic moments. Further, SC$_{\rm FP}$ persists over a wider field-angle range than typical of the compensation effect~\cite{Balicas2001}.




The temperature dependence of the SC$_{\rm FP}$ phase, as well as its close relation to the magnetic transition, is reminiscent of the field induced superconducting phase in URhGe, which has been attributed to ferromagnetic spin fluctuations associated with the competition of spin alignment between two weakly anisotropic axes. In URhGe, a magnetic field transverse to the direction of the ordered magnetic moments leads to the collapse of the Ising ferromagnetism and this instability enhances ferromagnetic fluctuations, which in turn induce superconductivity~\cite{Mineev2015}. 

UTe$_{2}$, however, is not ferromagnetic. Nevertheless, the overall similarities between UTe$_{2}$ and the ferromagnetic superconductors, with regard to the relationship between the preferred magnetic axis and the direction of high $H_{c2}$~\cite{Ran2019}, suggest that strong transverse spin fluctuations play a central role in these superconducting phases~\cite{Mineev2015}. The $H_{c2}$ values and directionality in UTe$_{2}$ can thus be understood in the following manner. Starting from zero magnetic field, superconductivity is most resilient to magnetic field applied along the $b$-axis, which is perpendicular to the easy magnetic $a$-axis. Magnetic field applied along the $b$-axis thus induces spin fluctuations that stabilize superconductivity against field-induced pair-breaking. At 34.5~T, however, a magnetic phase transition occurs, and magnetic moments rotate from the $a$ to the $b$-axis. In the high-field polarized phase, magnetic field along the $b$-axis no longer induces transverse spin fluctuations, and superconductivity is suppressed completely. However, it is possible once again to induce transverse spin fluctuations by now applying a magnetic field along the $c$-axis.  When viewed as a vector sum of fields along the $b$ and $c$-axes (Fig.~\ref{Reentrance}), it is clear that $H_b$ stabilizes the magnetic phase, while a range of $H_c$ strength stabilizes superconductivity, with the highest reentrant magnetic field values yet observed.  

This ferromagnetic fluctuation scenario is qualitatively consistent with the whole picture of field-induced superconducting phases in UTe$_{2}$, yet a very important distinction exists between the SC$_{\rm FP}$ phase and the field induced superconducting phase in URhGe: the SC$_{\rm FP}$ phase only exists in the field polarized state. This greatly challenges the current theory proposed for URhGe, which allows superconductivity to exist on both sides of the phase boundary~\cite{Mineev2015,Hattori2013,Sherkunov2018}.

A compelling resolution to this conundrum is that the SC$_{\rm FP}$ phase is the realization of a spin-triplet superconductor in the 1-dimensional quantum limit~\cite{Lebed2014}. This exotic superconductor requires spin-triplet pairing and is predicted to occur at very high magnetic fields applied transverse to the axis of a quasi-1-dimensional chain. The field-induced lower-dimensionality is both field-angle dependent and facilitates the recovery of the zero-field superconducting critical temperature, as we observe in UTe$_{2}$ (Figs.~\ref{Reentrance} and \ref{RESCT}).  Further, this mechanism permits superconductivity in a pure material to survive in any magnetic field, making UTe$_{2}$ an exciting playground for further testing the limits of high field-boosted superconductivity.

The exclusive existence of the SC$_{\rm FP}$ phase in the field polarized state, and in such high magnetic field, guarantees that the superconducting state has odd parity, with time reversal symmetry breaking. Odd parity is the cornerstone of topological superconductivity~\cite{Sato2017}, and it is certain that the SC$_{\rm FP}$ phase has non-trivial topology. Since time reversal symmetry is also broken, a special topological superconducting state, such as chiral superconductivity~\cite{Kallin2016}, is highly likely, which hosts Majorana zero modes, the building block for topological quantum computing~\cite{Sarma2015,Karzig2017}.

\section{methods}
Single crystals of UTe$_{2}$ were synthesized by the chemical vapor transport method using iodine as the transport agent. Crystal orientation was determined by Laue x-ray diffraction performed with a Photonic Science x-ray measurement system. Magnetoresistance measurements were performed at the National High Magnetic Field Laboratory (NHMFL), Tallahassee, using the 35-T DC magnet, and at the NHMFL, Los Alamos, using the 65-T short-pulse magnet. Proximity Detector Oscillator and magnetization measurements were performed at the NHMFL, Los Alamos, using the 65-T short-pulse magnet

In order to perfectly align the magnetic field along the $b$-axis in the DC magnet, a single crystal of UTe$_{2}$ was fixed to a home-made sample mount on top of an Attocube ANR31 piezo-actuated rotation platform (Fig.~\ref{2DanglePD}b). Thin, copper wires were fixed between the probe and rotation platform to measure the sample and two, orthogonal Toshiba THS118 hall sensors. All three measurements were performed using a conventional four-terminal transport setup with Lake Shore Cryotronics 372 AC resistance bridges. Adjustments to the $\theta$ angle were made using a low friction apparatus~\cite{Palm1999} to find the center of the range where the sample resistance was zero at $B$ = 25.5~T. With a lower field of 0.5~T, small changes were then made to the $\phi$ orientation while monitoring the Hall sensors. With the field aligned near to the $b$-axis, the magnetic field was swept to 34.5~T.

The contactless-conductivity was measured using the proximity detector oscillator (PDO) circuit described in Refs.~\cite{Altarawneh2009, Ghannadzadeh2011}, which has been used to study field stabilized superconducting phase~\cite{Singleton2000}. A coil comprising 6-8 turns of 46-gauge high-conductivity copper wire is wound about the single-crystal sample; the number of turns employed depends on the cross-sectional area of the sample, with a larger number of turns being necessary for smaller samples. The coil forms part of a PDO circuit resonating at 22-29 MHz. A change in the sample skin depth~\cite{Altarawneh2009}or differential susceptibility~\cite{Ghannadzadeh2011} causes a change in the inductance of the coil, which in turn alters the resonant frequency of the circuit. The signal from the PDO circuit is mixed down to about 2`MHz using a double heterodyne system~\cite{Altarawneh2009, Ghannadzadeh2011}. Data are recorded at 20 M samples/s, well above the Nyquist limit. Two samples in individual coils coupled to independent PDOs are measured simultaneously, using a single-axis, worm-driven, cryogenic goniometer to adjust their orientation in the field. 

The pulsed field magnetization experiments used a 1.5~mm bore, 1.5~mm long, 1500-turn compensated-coil susceptometer, constructed from 50 gauge high-purity copper wire~\cite{Goddard2008}. When a sample is within the coil, the signal is proportional to $dM/dt$. Numerical integration is used to evaluate magnetization M. The sample is mounted within a 1.3~mm diameter ampoule that can be moved in and out of the coil. Accurate values of M are obtained by subtracting empty coil data from that measured under identical conditions with the sample present. These results were calibrated against results from Quantum Design Magnetic Property Measurement System.

The data that support the results presented in this paper and other findings of this study are available from the corresponding authors upon reasonable request. Identification of commercial equipment does not imply recommendation or endorsement by NIST.

\section{addendum}
We acknowledge Fedor Balakirev for assistance with the experiments in the pulsed field facility. We also acknowledge helpful discussions with Andrei Lebed and Victor Yakovenko. Wesley T. Fuhrman is grateful for the support of the Schmidt Science Fellows program, in partnership with the Rhodes Trust. Research at the University of Maryland was supported by the US National Science Foundation (NSF) Division of Materials Research Award No. DMR-1610349, the US Department of Energy (DOE) Award No. DE-SC-0019154 (experimental investigations), and the Gordon and Betty Moore Foundation’s EPiQS Initiative through Grant No. GBMF4419 (materials synthesis). Research at the National High Magnetic Field Laboratory (NHMFL) was supported by NSF Cooperative Agreement DMR-1157490, the State of Florida, and the US DOE. A portion of this work was supported by the NHMFL User Collaboration Grant Program.   

[Competing Interests] The authors declare no competing interests.

[Correspondence] Correspondence and requests for materials should be addressed to S.R.~(email: sran@umd.edu) and N.P.B.~(nbutch@umd.edu).

\section{author contribution}
N. P. Butch directed the project. S. Ran, W. T. Fuhrman and S. R. Saha synthesized the single crystalline samples. S. Ran, I. Liu, and J. Singleton performed the magnetoresistance, PDO and magnetization measurements in pulsed field. Y. S. Eo, D. J. Campbell, P. Neves and D. Graf performed the magnetoresistance measurements in DC field. C. Eckberg and H. Kim performed magnetoresistance measurements in low magnetic fields. S. Ran and N. P. Butch wrote the manuscript with contributions from all authors. 

\bibliography{UTe2HF}

\clearpage

\clearpage

\section*{Supplementary Text}
\subsection{SC$_{RE}$ phase in $bc$ plane}
\begin{figure}
\includegraphics[angle=0,width=120mm]{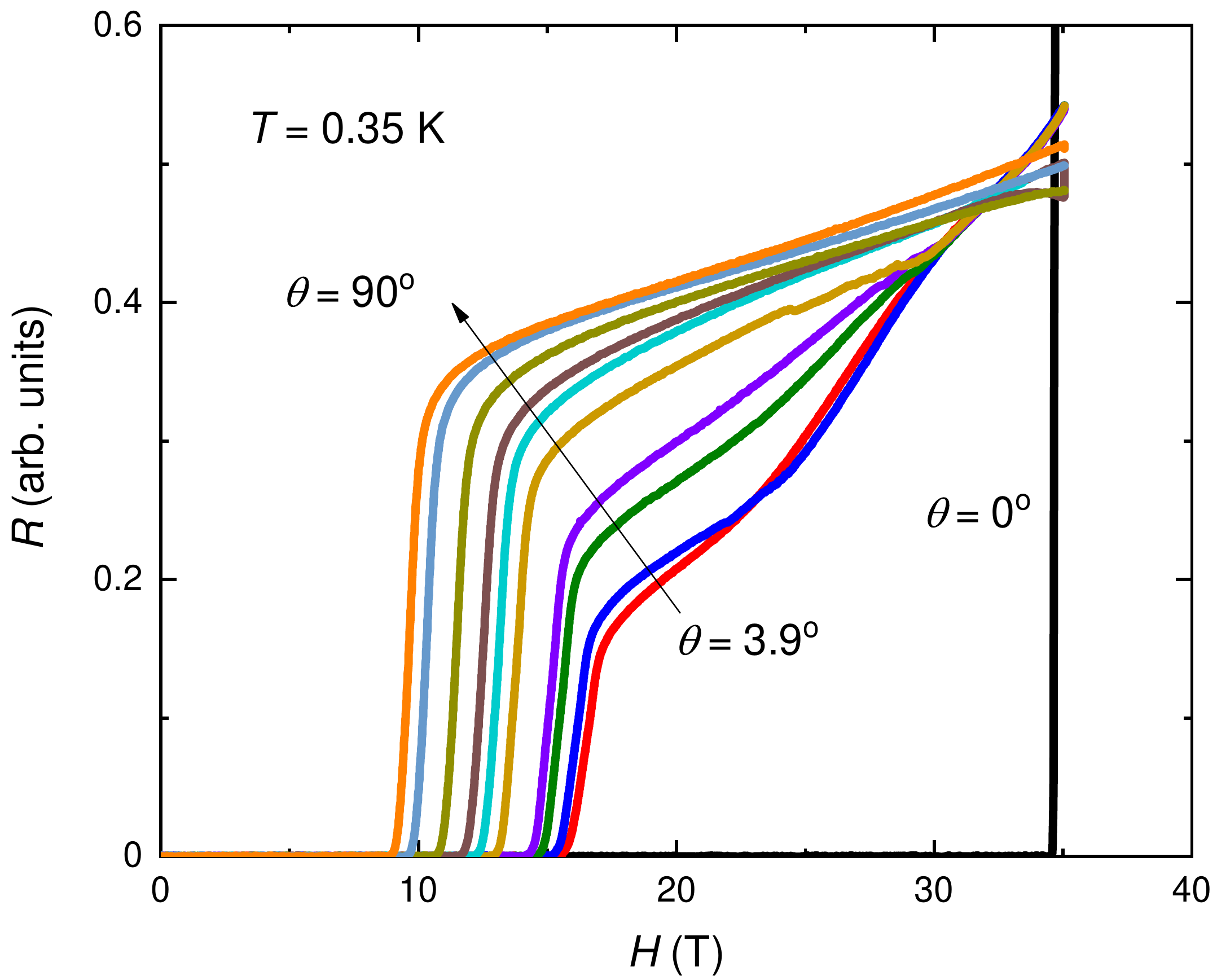}
\caption{Field dependence of magnetoresistance of UTe$_{2}$ at $T$ = 0.35~K measured in the DC field. The magnetic field is rotated from $b$ towards $c$ axis. Zero resistance persists up to 34.5~T when the magnetic field is perfectly along $b$-axis. Reentrance of superconductivity is not observed when the magnetic field is rotated from $b$ towards the $c$-axis.} 
\label{bcangle}
\end{figure}

We did not observe the SC$_{RE}$ phase for $\theta$ larger than 3.9$^{\circ}$ in the $bc$-plane. It is very likely that this is the angle limit of the SC$_{RE}$ phase. The magnetoresistance data show a slope change between 20 and 30~T for $\theta$ = 3.9$^{\circ}$ (Fig.~\ref{bcangle}), indicating it is on the edge of SC$_{RE}$ phase. 

\subsection{Angle offest in pulsed field measurements}
In order to detect the magnetic transition at higher angles, we had to perform experiments in pulsed field, where a probe compatible with a two-axis rotator is not available. Therefore, when the magnetic field rotates in one plane (e.g., $ab$-plane), there generally is a small angle offset in the perpendicular plane (e.g., $bc$-plane). This is probably why the field range of SC$_{PM}$ looks smaller in pulsed field compared to the value in our previous paper, and SC$_{RE}$ is not observed, in both magnetoresistance and PDO measurements. In addition, the base temperature in pulsed field, 0.5~K, is higher than that of the DC field, 0.35~K. For these reasons, pulsed field data are not used to characterize the SC$_{PM}$ and SC$_{RE}$ phases. 

The field polarized state and SC$_{FP}$ phase extends to very high fields, beyond the limit of DC field, and therefore the pulsed field measurements are the only choice to characterize both phases, which inevitably gives some angle offset. This angle offset, of order a few degrees, explains the slight difference between the phase diagrams based on PDO and magnetoresistance measurements. 

\subsection{hysteresis in PDO and magnetoresistance}
The high field induced superconducting phase SC$_{FP}$ can be seen in both PDO and magnetoresistance measurements upon down-sweep of the magnetic field. PDO measurements show a sudden increase of frequency, corresponding to the decrease of the sample resistance. However, the frequency decreases instead upon the up-sweep of the magnetic field, leading to an unusual, large hysteresis loop between up and down sweep. Similarly, magnetoresistance shows a sudden increase upon the up-sweep, leading to large hysteresis. On the other hand, the hysteresis in magnetoresistance decreases with increasing angle from $b$ towars $c$-axis and almost disappears for $\theta$ = 24$^{\circ}$. 

Most of the hysteresis is known to be caused by heating due to dissipative vortex motion in the mixed state during the rapid up-sweep of the field. The rise time to peak field upon the up-sweep is about 9~ms, and the down-sweep time is about 90~ms, hence $dB/dt$ is much larger as the field increases. Therefore, the sample is relatively hot upon up-sweep as it exits the vortex state. Based on the temperature dependence of the SC$_{FP}$ phase, it turns into field polarized state above the critical temperature. In such a scenario, what we observed upon up-sweep and down-sweep corresponds to the magnetic transition and superconducting reentrance, respectively. However, once the samples are in the normal state, there is known to be little heating due to the changing field. This is consistent with the fact that the hysteresis is only observed for the superconducting phase, not the field polarized phase.

Sample sizes are kept (i) small to present very little cross-sectional area to the field (thereby minimizing eddy-current heating) and (ii) thin to provide a large surface area-to-volume ratio to maximize cooling. In addition, rapid thermalization is assisted by using a relatively high pressure of $^4$He exchange gas.
Finally, as mentioned above, during the down-sweep, $dB/dt$ is significantly smaller than during the up-sweep, further reducing any
residual eddy-current heating. Hence, the sample is essentially in equilibrium with the thermometer when it enters the vortex
state on the way down, leading to an accurate measurement of the transition. Therefore, we use the down-sweep data for determining the phase diagrams.

There are other possible sources for the hysteresis. It might be that the SC$_{FP}$ phase is unusual, and that the magnetic susceptibility, rather than the conductivity, dominates the PDO response. The hysteresis loop we observed can be due to the irreversibility field of superconductivity. Also, the vortices in the SC$_{FP}$ phase can cause high-frequency dissipation during the up-sweep, which is distinct from the heating mentioned above. Finally, the spatial distribution of the screening currents within the sample could be different on the up-sweep and down-sweep. This could affect the sample’s ability to screen radio frequency fields, leading to hysteresis in the PDO measurement. 


In PDO measurements, we also observed hysteresis for $\theta$ = 0$^{\circ}$. It is not clear whether the increase in frequency upon the up-sweep is related to the SC$_{RE}$ phase or the magnetic transition. 

\subsection{criteria to determine the critical fields}
In order to construct the phase diagram with consistent critical field values, the following criteria are used to extrapolate the critical fields of various phases: for magnetoresistance measurements, we use the field at which the maximum slope of the resistance data that goes to zero resistance extrapolates to zero resistance (Fig.~\ref{criteria}a and b); for PDO measurements, we use the field at which the maximum in derivative occurs (Fig.~\ref{criteria}c and d).  

\subsection{Magnetization measurements}
The magnetization measurement was performed with the magnetic field applied at $\theta$ = 35$^{\circ}$ from $b$ towards $c$-axis, where the SC$_{FP}$ phase was observed in magnetoresistance and PDO measurements. Similar to what seen for field along $b$-axis, the magnetic moment jumps from 0.4 to 0.7~$\mu_{B}$, indicating a field polarized state in the high magnetic field. 

\clearpage

\begin{figure}
\includegraphics[angle=0,width=150mm]{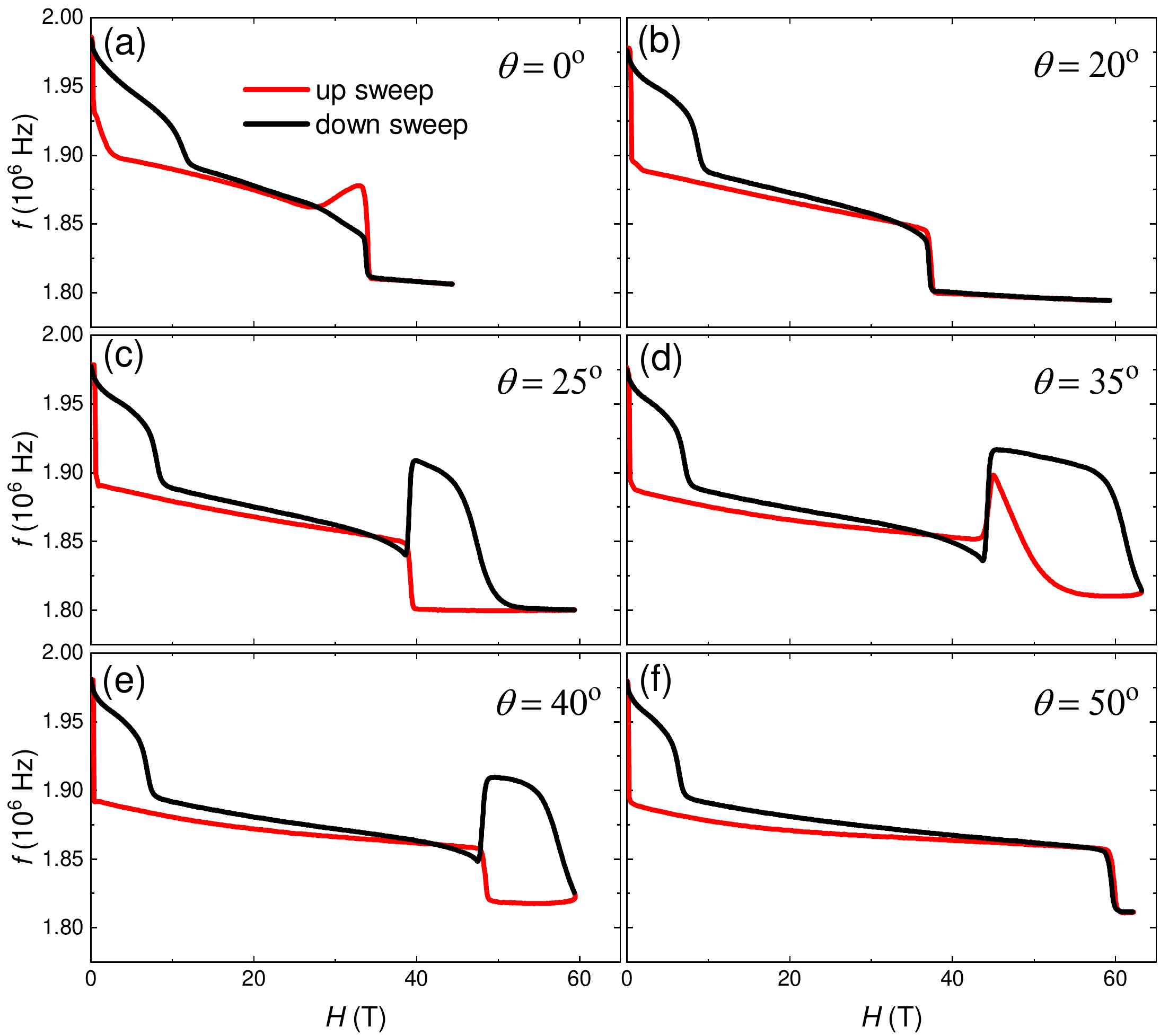}
\caption{PDO measurements of UTe$_{2}$ at $T$ = 0.45~K in the  pulsed  field, for magnetic field applied at various angles from $b$ towards the $c$-axis}
\label{PDOhys}
\end{figure}

\begin{figure}
\includegraphics[angle=0,width=150mm]{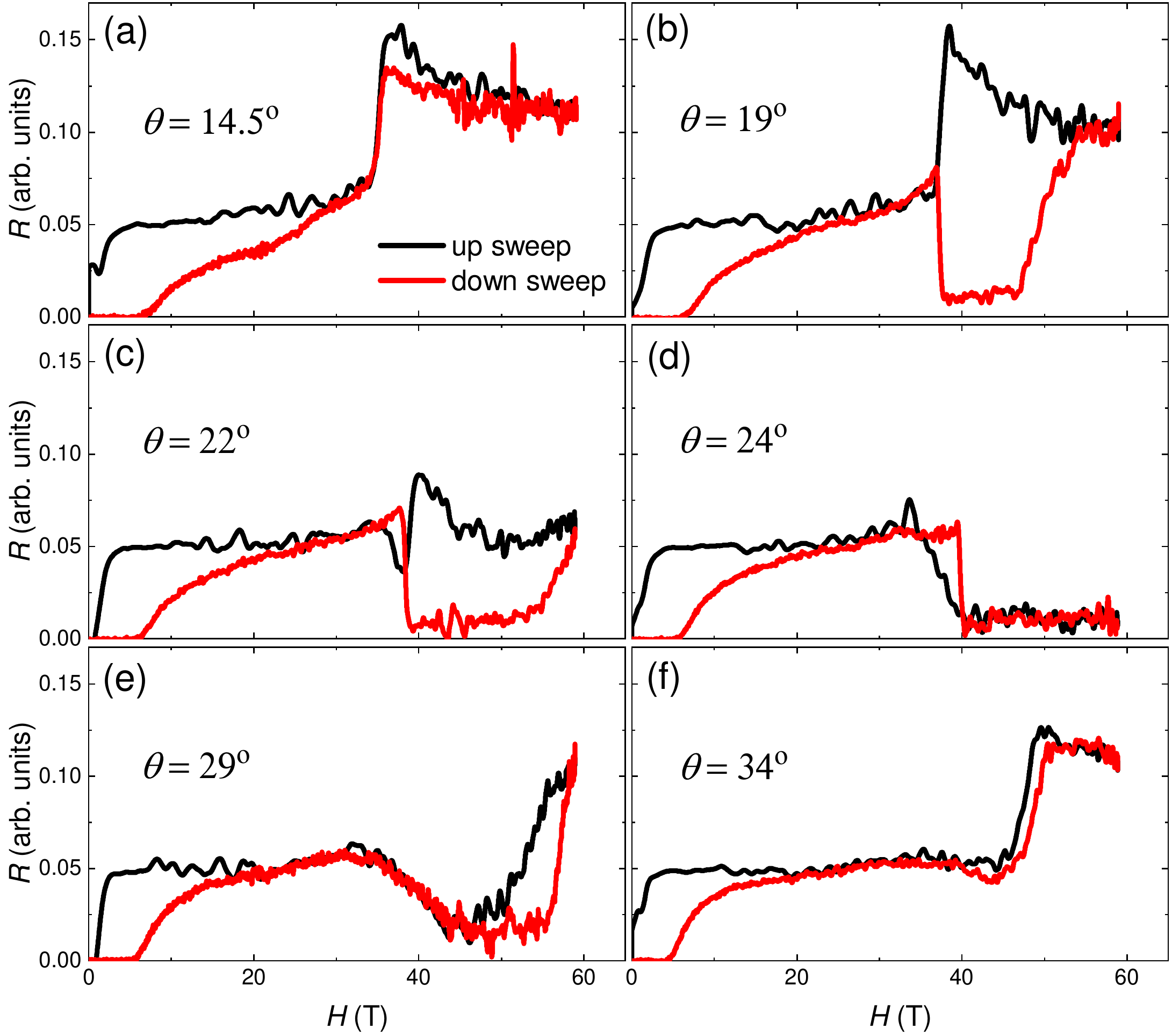}
\caption{Magnetoresistance measurements of UTe$_{2}$ at $T$ = 0.45~K in  the  pulsed  field, for magnetic field applied at various angles from $b$ towards the $c$-axis.}
\label{MRhys}
\end{figure}

\begin{figure}
\includegraphics[angle=0,width=150mm]{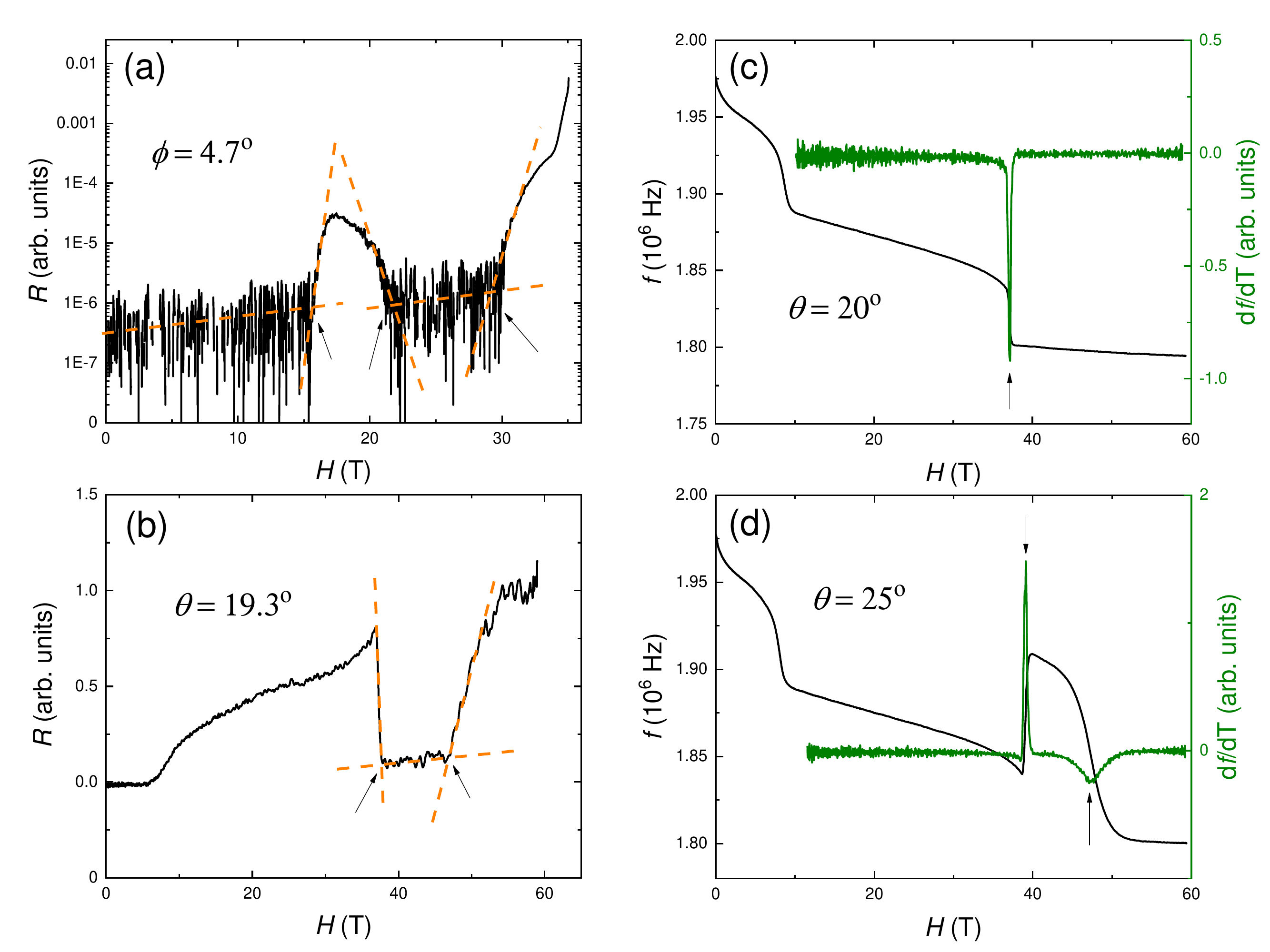}
\caption{Selected magnetoresistance (a nd b) and PDO (c and d) measurements to show the criteria used to extrapolate the critical field values for various phases.}
\label{criteria}
\end{figure}

\begin{figure}
\includegraphics[angle=0,width=120mm]{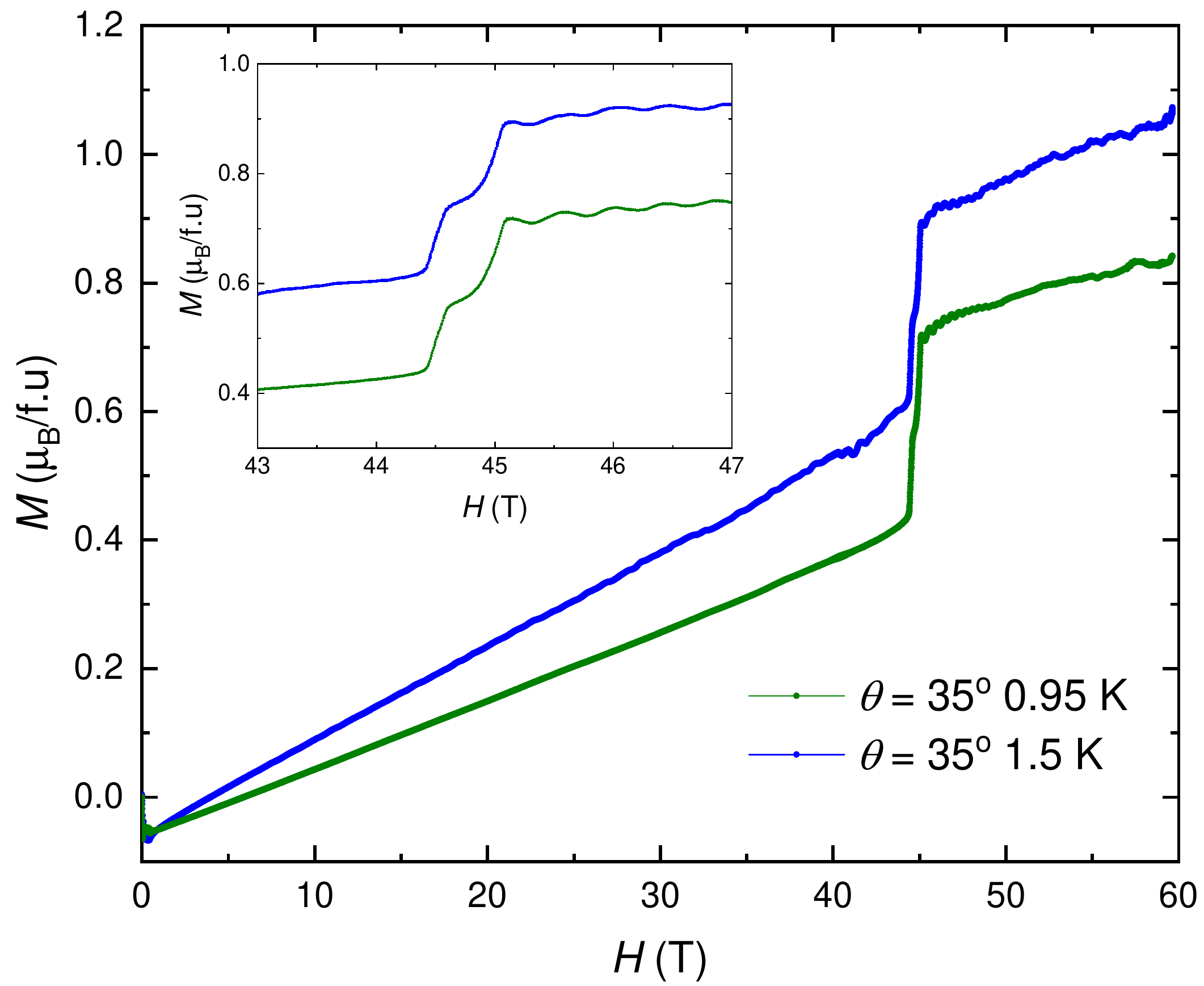}
\caption{Magnetization measurements UTe$_{2}$ at $T$ = 0.95~K and 1.5~K in the pulsed field, with the magnetic field applied at $\theta$ = 35$^{\circ}$ from $b$ towards $c$-axis.}
\label{M35deg}
\end{figure}

\end{document}